\documentstyle[psfig,amssymb]{mn}

\title[A triple nucleus in the BCG in Abell 193]
{A triple nucleus in the Brightest Cluster Galaxy in Abell 193$^*$}

\author[Seigar et al.] 
{Marc S. Seigar$^{1, 2}$,  Paul D. Lynam$^3$ and Nicole E. Chorney$^{1, 4}$\\ 
$^1$Joint Astronomy Centre, 660 N. A'ohoku Place, Hilo, HI 96720, USA\\
$^2${\tt email: m.seigar@jach.hawaii.edu}\\
$^3$Max-Planck-Institute for Extraterrestrial Physics, D-85748 
Garching bei M\"unchen, Germany\\
$^4$Department of Physics \& Astronomy, University of Victoria, PO Box 3055 
STN CSC, Victoria BC, V8W 3P6, Canada\\
$^*$Based, in part, on observations made with the NASA/ESA 
Hubble Space Telescope, obtained from the archive at the Space\\
Telescope Science Institute. STScI is operated by the Association 
of Universities for Research in Astronomy, Inc. under NASA\\
contract NAS 5-25555.\\
}

\begin{document} 
\maketitle

\begin{abstract}
We present a ground-based near-infrared {\em K} band image and an HST/WFPC2 
image of the brightest cluster galaxy in Abell 193 (IC 1695). This object was
selected as the central cluster galaxy using X-ray information. Both images
reveal a triple nucleus structure. Previously, this galaxy was thought to 
have only 2 nuclei. We present colours and magnitudes and a colour plot of 
the three nuclei. The nuclear structure and colours of the nuclei in this
galaxy suggest that a merger may have taken place in its recent history.
\end{abstract}
  

\section{INTRODUCTION}

Our picture of the central structure of galaxies has evolved significantly
over the last decade, due to the high resolution capabilities of the
{\em Hubble Space Telescope} (HST) and ground-based facilities with active and
adaptive optics. Optical observations using WFPC2 have enabled detailed
analysis of the cores of elliptical (Lauer et al. 1995; Byun et al. 1996;
Gebhardt et al. 1996; Faber et al. 1997; Graham et al. 2003) and 
spiral (Carollo et al. 1997; Carollo, Stiavelli \& Mack
1998; Carollo \& Stiavelli 1998) galaxies, with extension to near-infrared
wavelengths using NICMOS (Carollo et al. 2001, 2002; Seigar et al. 2000,
2002; Balcells et al. 2003; Fathi \& Peletier 2003).

Nuclear investigations of elliptical galaxies, to date, have 
concentrated on normal ellipticals with no attempt to understand 
properties of giant ellipticals. Of a sample of 57 elliptical galaxies,
Lauer et al. (1995) included only 6 giant elliptical galaxies and say
little about their properties. A more recent study by Laine et al. (2003)
used HST to study a sample of 81 giant elliptical galaxies to study their
nuclear morphologies. The HST image presented in this paper was obtained as
part of the the snapshot program on which the Laine et al. (2003) study is 
based. The nuclei of giant elliptical galaxies 
are of interest, because these galaxies tend to lie at the centres of
clusters as the brightest cluster galaxy (BCG).
About half of all BCGs have companions, or
a secondary nucleus, located within a projected radius of 10 $h^{-1}$ kpc
($H_{0}=100 h$ kms$^{-1}$Mpc$^{-1}$) of the luminosity centre of the combined
system (Hoessel \& Schneider 1985). Chance projections of background or
foreground cluster galaxies into this radius are expected to occur only about
10\% of the time for a cluster of richness class 2 (Tonry 1984). Thus there is
a real excess of galaxies near first-ranked cluster members; no similar 
excess exists around the second or third ranked galaxies (Schneider, Gunn \&
Hoessel 1983). Abell 193
has a richness class of 1, and therefore the chance of foreground or
background projections will be slightly lower.

These ``multiple nuclei'' were at one time taken to be the best evidence
for models which postulate that massive galaxies are growing at the centres of
rich clusters by accreting, or ``cannibalizing'' their less massive neighbours
(Hausmann \& Ostriker 1978). The major difficulty with this interpretation is
that the radial velocities of the secondary concentrations 
relative to that associated with the centre of the underlying BCG light
distribution are typical
of randomly selected cluster galaxies (Jenner 1974; Tonry 1984, 1985). These
high velocity differences imply that the ``nuclei'' are generally not bound,
and are, therefore, not candidates for imminent cannibalism (Merritt 1984).

In this letter we report the discovery of a triple nucleus in the centre of
IC 1695, the BCG in Abell 193. IC 1695 was identified as the BCG in Abell
193, optically by Lauer \& Postman (1994), and using X-rays by 
Lynam et al. (2000).
This galaxy was previously only thought to
contain two nuclei (Hoessel \& Schneider 1985). We present two images,
revealing a triple nucleus, one in the near-infrared {\em K} band, taken
from the ground with the 3.8-m UK Infrared Telescope (UKIRT)
on Mauna Kea, Hawaii, and 
the other taken with WFPC2 on HST with the F814W (i.e. {\em I} band) filter.
We also present spectra taken with UKIRT. The spectra were used to calculate
the redshifts of the three central components.
We then discuss the possibility of this being the result of a merging system.

This letter is arranged as follows. In Section 2 we report the observations;
Section 3 is a discussion of the results of these observations; Section 4 is
a discussion of the interpretation of nuclear characteristics in galaxies; 
Section 5 summarises our conclusions.

\section{OBSERVATIONS}

We have obtained a near-infrared {\em K} band (2.2 $\mu$m) image of the BCG
in Abell 193 (IC 1695). This image was observed at the 3.8-m 
UK Infrared Telescope
(UKIRT) with the UKIRT Fast Track Imager (UFTI) on 30 July 2002. 
UFTI is a 1024$\times$1024 Hawaii array with a pixel scale of 
0.\hspace*{-1mm}$^{\prime\prime}$091.
A total
of 9 frames were observed with an individual exposure time of 60 seconds. A
9-point jitter pattern was used with offsets of $\sim$ 20 arcsec between
each frame, in order to facilitate cosmic ray and bad pixel removal. 
After dark subtraction, each frame was flat-fielded using interspersed sky
frames of equal integration time to the science frames. The flat-fielded 
frames were then
masked for bad pixels and sky subtracted. The final frames were then combined 
to provide a single image of effective integration time 9 minutes.
At the time when this image was taken, the seeing was extremely good with
FWHM$\simeq$0.\hspace*{-1mm}$^{\prime\prime}$35.

We have also retrieved a {\tt WFPC2} image of this galaxy from the 
{\em Hubble Space Telescope} (HST) archive. The image was taken with the 
{\tt F814W} ({\em I} band) broadband filter and was centred on the {\tt PC}
with a pixel scale of 0.\hspace*{-1mm}$^{\prime\prime}$046. 
A total of 2 frames were observed
and combined in order to remove cosmic rays. The exposure time of the combined
image was 1000 seconds.

We have also obtained $K$ band spectra of the three nuclei using CGS4 on UKIRT.
The spectra were obtained on 4 December 2002. The integration times were scaled
so as to get approximately the same signal-to-noise ratio for each nucleus.
The integration time on the brightest nucleus was 60 minutes, on the next
nucleus 90 minutes and the final nucleus was observed for 100 minutes. The slit
was oriented E-W, such that light from only one nucleus fell in the beam. The
central wavelength was 2.2$\mu$m and the 40 line/mm grating was used with a 0.6
arcsec slit. This resulted in a resolving power of $\sim$800.

\section{RESULTS}

Figure 1 shows the inner 16$^{\prime\prime}\times$16$^{\prime\prime}$ of
both the
{\em K} band image observed at UKIRT (left panel)
and the {\em I} band image observed with HST (right panel). Both clearly
show the triple nucleus structure in the centre of IC 1695.

\begin{figure*}
\label{ic1695}
\includegraphics{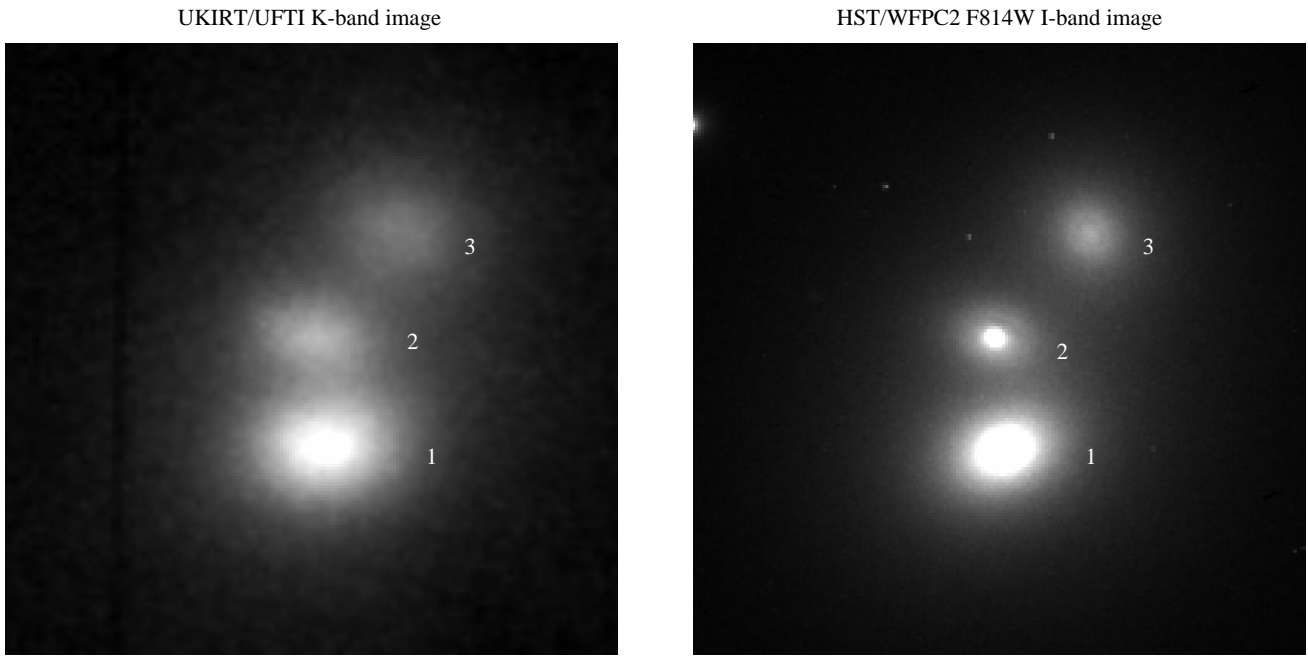}
\vspace*{9cm}
\caption{Left panel: UKIRT/UFTI {\em K} band image of the centre of IC 1695. Right panel: HST/WFPC2 F814W {\em I} band image of the centre of IC 1695. Both images show the triple nucleus in the centre of this galaxy.}
\end{figure*}

Using these images we have calculated the brightness of the three 
nuclei using three different methods. In the first of these we
fitted a Gaussian to each of the nuclei, setting the background
level to that of the underlying galaxy. In the second method, we
fitted a polynomial to the underlying galaxy and subtracted the
resulting fit. The total counts in each nucleus were then summed,
using fixed apertures, designed to encompass the light from only
one of the nuclei. Elliptical apertures were used with semi-major
axis, $a=2.$\hspace*{-1mm}$^{\prime\prime}00$ for nucleus 1, 
$a=1.$\hspace*{-1mm}$^{\prime\prime}00$ 
for nucleus 2 and $a=1.$\hspace*{-1mm}$^{\prime\prime}73$ for nucleus 3.
For the final method an iteratively smoothed and nuclei-masked 
version of the original frame was subtracted from the latter,
and the measurement was performed on the resulting image 
(containing only the nuclei). Again the counts in each nuclei were
summed within the same fixed apertures as in method one. 
The three methods are complementary; 
they sample the galactic background in different regions, and 
hence, their combined use allows us to estimate the uncertainty in 
the derived values contributed by the poorly constrained underlying
galactic light. Each method gives a very similar result. 
We use the mean of the three estimates as our
final measurement of the brightness of the nuclei. The error on this
mean is taken as a standard error in the three measurements. 
The resulting brightnesses and
colours of the nuclei are shown in Table 1. Photometric calibration of the HST
image was performed using the measured WFPC2 zero-points reported in Holtzman 
et al. (1995). For the UFTI {\em K} band 
image, UKIRT faint standards were observed
throughout the night, in order to estimate the zero-point and extinction
correction, using the standard star magnitudes reported in Hawarden et al. 
(2001).

   \begin{table}
      \caption{Magnitudes and colours of the three nuclei. See Figure 1 for the numbering scheme used to identify the nuclei}
         \label{colours}
	\begin{center}
         \begin{tabular}{llll}
            \hline
	    \hline
	    \noalign{\smallskip}
            Nucleus	& {\em K} band		& {\em I} band		& $I-K$	\\
            \noalign{\smallskip}
            \hline
            \noalign{\smallskip}
	    1		& 14.58$\pm$0.05 	& 16.32$\pm$0.02	& 1.74$\pm$0.05\\
	    2		& 15.77$\pm$0.10 	& 17.12$\pm$0.03	& 1.35$\pm$0.10\\
	    3		& 15.91$\pm$0.13 	& 17.48$\pm$0.06	& 1.57$\pm$0.14\\
            \noalign{\smallskip}
            \hline
         \end{tabular}
	\end{center}
   \end{table}

Table 2 shows the separation of the nuclei. This is based upon a measured
heliocentric recessional velocity of 14505 kms$^{-1}$ (de Vaucouleurs et al. 
1991) for IC 1695 and an 
assumed Hubble constant of $H_{0}=75$ kms$^{-1}$Mpc$^{-1}$. From this the
distance to IC 1695 is 193.4 Mpc. The precision with which one can measure
the angular distances is assumed to be no better than one pixel. Both the HST 
image and the UKIRT image have been used for estimating the separations 
between the nuclei. The uncertainties are lower with the HST image, due to
the higher resolution and the smaller pixel scale of WFPC2. The
separations measured with the UKIRT image demonstrate that the measured
distances are consistent with each other.

   \begin{table}
      \caption{Distance between the nuclei. Column 1 reports the 2 nuclei for which the distance has been measured. Columns 2 and 3 have been calculated using the HST/WFPC2 F814W image. Columns 4 and 5 have been calculated using the UKIRT/UFTI K-band image.}
         \label{distances}
	\begin{center}
         \begin{tabular}{lllll}
            \hline
	    \hline
	    \noalign{\smallskip}
	    Nuclei	& \multicolumn{2}{c|}{HST F814W image}		& \multicolumn{2}{c|}{UKIRT K-band image}	\\
            		& Distance		& Distance		& Distance		& Distance		\\
			& (arcsec)		& (kpc)			& (arcsec)		& (kpc)			\\
            \noalign{\smallskip}
            \hline
            \noalign{\smallskip}
	    1 -- 2	& 1.70$\pm$0.05		& 1.60$\pm$0.04		& 1.66$\pm$0.09		& 1.56$\pm$0.09		\\
	    1 -- 3	& 3.47$\pm$0.05		& 3.25$\pm$0.04		& 3.43$\pm$0.09		& 3.22$\pm$0.09		\\
	    2 -- 3	& 2.08$\pm$0.05		& 1.90$\pm$0.04		& 2.02$\pm$0.09		& 1.89$\pm$0.09		\\
            \noalign{\smallskip}
            \hline
         \end{tabular}
	\end{center}
   \end{table}

From the distances calculated between the nuclei it is possible for us to
determine which of these nuclei had not been uncovered in earlier work.
Hoessel \& Schneider (1985) identified IC 1695 as having a double nucleus,
and calculated a distance of $\sim$3.\hspace*{-1mm}$^{\prime\prime}$5
between them. It therefore seems that we have been able to determine that
nucleus 1 and 2 are separate features for the first time, whereas earlier
observations did not have the capabilities to resolve these two features.

Figure 2 shows a colour image of the nuclei in IC 1695. This has been 
carefully aligned by sub-dividing each pixel in the UFTI array (using the {\em
magnify} routine in {\tt IRAF}) into a 2$\times$2 square, resulting in
approximately the same pixel scale as the PC on WFPC2. The orginal {\em K} 
band image was convolved with the HST point spread function (psf) and the HST 
image was convolved with the psf of the UKIRT {\em K} band image. The centre 
of each of the nuclei was calculated by fitting a Gaussian, and they were then 
aligned. The pixels were then binned up to the orginal UFTI pixel scale. The 
images have therefore been aligned to within 1 WFPC2 pixel.

\begin{figure}
\label{colour}
\includegraphics{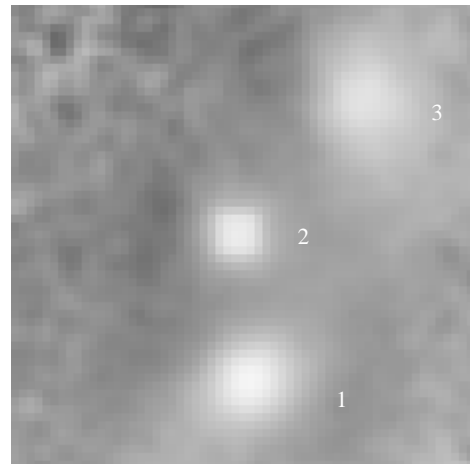}
\vspace*{6cm}
\caption{$I-K$ colour picture of the nuclei in IC 1695.}
\end{figure}

The colour image shown in Figure 2 suggests that the nuclei are bluer in colour
than the typical colour of a BCG at $z=0.048$, which is $I-K\simeq1.9$. The
colours of the nuclei (listed in table 1) have been calculated pixel by pixel
from figure 2. The bluer colour suggests that younger stars exist in this
system than in a typical BCG. This could be due to a merger that may have
taken place in the recent history of this galaxy. This would 
also account for the ``multiple-nuclei'' structure.
If the triple nucleus structure is the result of a merger
between three galaxies, it would be expected that shortly after the merger, a
burst of star formation was triggered, due to the large tidal forces involved
(e.g. Cui et al. 2001). A possible result of such a 
phase in the recent history of a galaxy would be 
to make it appear bluer than other
BCGs at the same redshift, due to the formation of a younger stellar 
population.

We have also used $K$ band 
spectra of each nucleus to estimate its redshift, and 
therefore we can say with much more certainty that all three nuclei are
part of the same system, rather than just projections of foreground or
background galaxies. The spectra are shown in Figure 3. They are devoid
of emission lines, suggesting that this galaxy has no (or very little) gas.
However, the redshift can be calculated from the 2.29$\mu$m CO absorption 
feature. This has been modelled using stellar template spectra and calculating
where this feature begins.

\begin{figure*}
\label{nuc}
\includegraphics{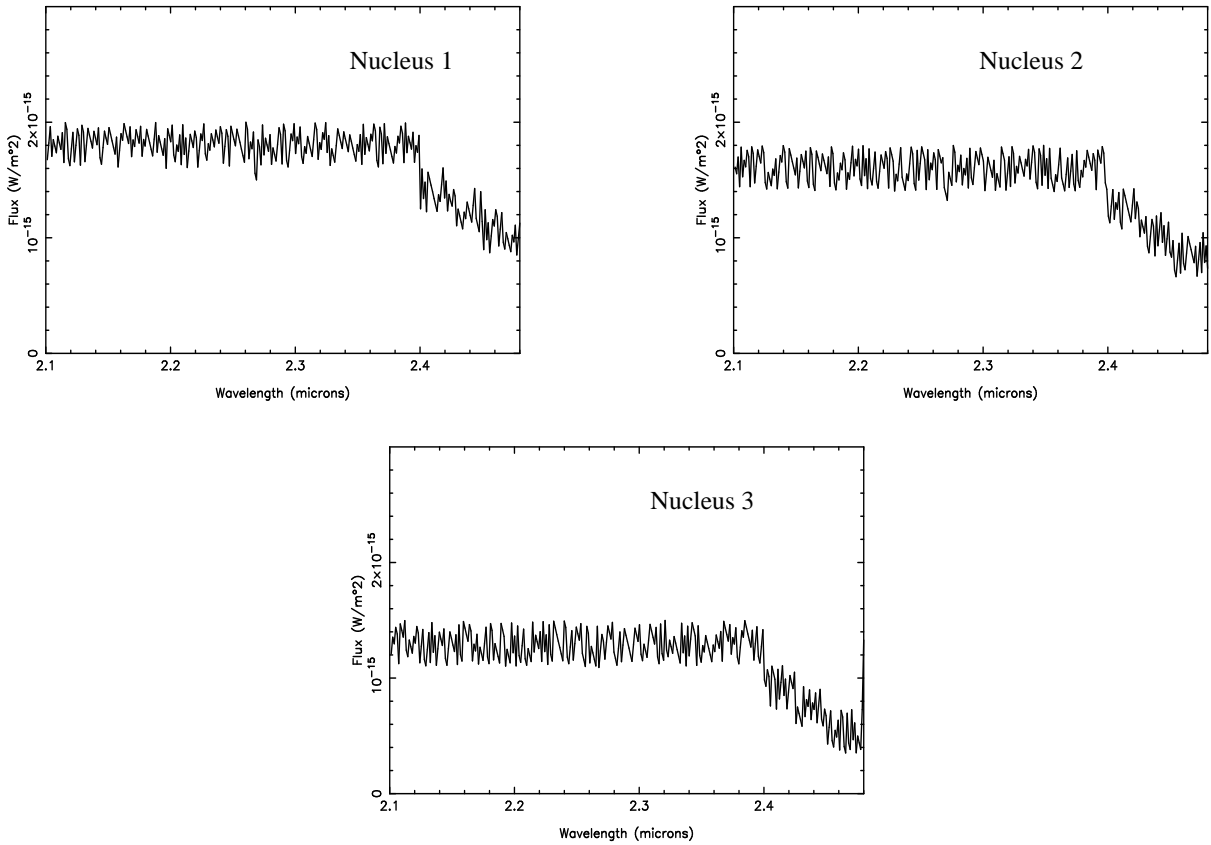}
\vspace*{11.7cm}
\caption{$K$ band spectra of the 3 nuclei, showing the 2.29$\mu$m CO
absorption feature redshift to $\sim$2.4$\mu$m.}
\end{figure*}

From the spectra the redshift and recessional velocity of each nucleus has 
been calculated. As
can be seen, the CO feature is redshifted to $\sim2.4\mu$m in each nucleus,
and this corresponds to a redshift of $z\simeq0.048\pm0.001$. In this case
the error has been calculated from the resolving power available with CGS4
and the 40 line/mm grating using a 0.6 arcsec slit. The $S/N$ is also taken
into account in the error. This redshift corresponds to a
recessional velocity of 14400$\pm$300 kms$^{-1}$. The error in the 
recessional velocity is similar to the typical velocity dispersion of an
elliptical galaxy.  
It would therefore be difficult to achieve a better estimate
of the recessional velocities. As a result, we can say that within the errors,
the three nuclei all lie at the same redshift.

This result, combined with the fact that the nuclei are bluer than expected
for a typical BCG, seems to suggest that the three nuclei are gravitationally
bound within the same system. 

\section{DISCUSSION}

As already discussed the chance projection of a background or foreground 
galaxy falling within a projected radius of 10 $h^{-1}$kpc 
($H_0$=100$h$ kms$^{-1}$Mpc$^{-1}$) is 10\% (Tonry 1984; Hoessel \& Schneider 
1985). In otherwords, if IC 1695 actually has a double--nucleus, then the 
chance of the third nucleus being a projection of a foreground or background
galaxy is 10\%. However, if only one of these nuclei belongs to IC 1695, and
both other nuclei are projections, then the chance of this is only 1\%, and 
therefore unlikely, 
unless one observes a sample containing hundreds of brightest cluster galaxies.
The fact that clusters can be slightly elliptical, with their major axes
being aligned with the major axes of giant elliptical galaxies 
(Porter, Schneider \&
Hoessel 1991) may slightly increase this chance, depending on the line of
sight we have with respect to the major axis of the cluster. 
However, it still remains a slim
chance. 

Taking into account the blue colours of the nuclei, this suggests that
all three galaxies would have actively been forming a young population
of stars in their recent histories. This is unlikely for elliptical galaxies,
unless they have been interacting in some way. 
Finally, the calculated redshifts for
all three nuclei puts them at the same redshift, with an error consistent with
the typical velocity dispersion of an elliptical galaxy. The simplest 
conclusion is that the three nuclei are gravitationally bound as parts of a 
common merged, or merging system.

Although this system may have recently been forming stars, this activity now
seems to have ceased, and the galaxy spectra show no emission lines, due
to very little (or no) gas being present to fuel further star formation.
Optical spectra observed by Owen et al. (1995) have also showed that there
is no H$\alpha$ emission from this galaxy, and this confirms the absence of
any hydrogen gas. However, Owen et al. (1995) did find an absolute line 
luminosity of $3\times10^{32}$ W for the [O{\tt II}]3727 A line. At the 
distance of IC 1695, this corresponds to an apparent line strength of 
$1.20\times10^{-16}$ W/m$^2$. This line is probably present, 
due to the fact that IC 1695 is a weak radio galaxy (Owen et al. 1995).

The timescale on which the three components 
will merge depends on several factors.
The density and velocity dispersion of a cluster determines the frequency at
which mergers occur. The centres of clusters are the most dense areas and so
the likelihood of finding a BCG with multiple nuclei is significant. In fact,
Ryden et al. (1993) found that over 25\% of giant ellipticals in their sample
had multiple nuclei. To find BCGs with multiple nuclei is therefore quite 
common, and one would assume that the timescale on which the multiple nuclei
merge is of the order of 10$^9$ years (Murphy et al. 2001). This timescale
has been calculated by studying the merging
sequence of ULIRGs, via imaging spectroscopy of the Pa$\alpha$
line. This data has shown that even after 10$^9$ years, distinct nuclei
can be still be seen in merging systems. Such a timescale could 
explain why multiple nuclei are seen in so many central cluster galaxies.

Having said this it is important to note that the most conducive place for
mergers is in the centres of intermediate density clusters and large groups,
not necessarily the richest clusters (B\"ohringer, private communication). 
Density is important, but in the 
richest, most massive clusters the interaction velocities are so high
that the galaxies can pass right through and by each other, but their
relative velocity is high enough that they never become gravitationally
bound to each other, and therefore a common, merged system cannot form.

\section{CONCLUSIONS}

We have observed a {\em K} band image of the BCG in Abell 193 (IC 1695) in
exceptional seeing conditions 
(FWHM$\sim$0.\hspace*{-1mm}$^{\prime\prime}$35). This image 
has revealed a triple nucleus structure, in a galaxy which was formerly
thought to 
contain only two nuclei. This structure has been confirmed with the use of
an HST archive WPC2 F814W I-band image. The $I-K$ colours 
revealed by the images
suggest that the nuclei are bluer than expected from typical BCG colours at
similar redshifts. This combined with the spectra, that place the three nuclei
at the same redshift of $z\simeq0.048\pm0.001$, suggests that the three nuclei
are indeed part of the same system.

\section{ACKNOWLEDGMENTS}

The United Kingdom Infrared Telescope is operated by the Joint Astronomy 
Centre on behalf of the U.K. Particle Physics and Astronomy Research Council.
This research has made use of the NASA/IPAC Extragalactic Database (NED) which 
is operated by the Jet Propulsion Laboratory, California Institute of 
Technology, under contract with the National Aeronautics and Space 
Administration. The authors wish to thank the anonymous referee for suggestions
which improved the content of this paper.

\end{document}